\documentclass[twocolumn]{article}
\usepackage{graphicx} % Required for inserting images
\usepackage{tabularx}
\usepackage{rotating}
\usepackage{nameref}
\usepackage{array}
\usepackage{standalone}
\usepackage{comment}
\usepackage[hyphens]{url} % Allow line breaks at hyphens
\usepackage{hyperref}  

\usepackage{makecell}
\usepackage{listings}
\usepackage{xcolor}
\usepackage{caption}
\usepackage{listingsutf8} 
\usepackage{float}       
\usepackage{etoolbox}     
\usepackage{newfloat}     
\usepackage{chngcntr}
\usepackage{authblk}
\usepackage{orcidlink}
\usepackage[labelfont=bf]{caption}
\usepackage[most]{tcolorbox}
\tcbuselibrary{listings, breakable}

\DeclareCaptionType{listing}[Listing][List of Listings]

\hypersetup{
    colorlinks=true,
    linkcolor=blue,
    filecolor=magenta,      
    urlcolor=blue
    }

\title{Identity resolution of software metadata using Large Language Models}
\date{}

\author[1]{Eva Martín del Pico\orcidlink{0000-0001-8324-2897}\thanks{Corresponding author: eva.martin@bsc.es}}
\author[1,2]{Josep Lluís Gelpí\orcidlink{0000-0002-0566-7723}}
\author[1]{Salvador Capella-Gutiérrez\orcidlink{0000-0002-0309-604X}}

\affil[1]{Barcelona Supercomputing Center, Barcelona, Spain}
\affil[2]{Department of Biochemistry and Molecular Biology, University of Barcelona, Barcelona, Spain}

\begin{document}

\maketitle

\begin{abstract}
Software is an essential component of research. However, little attention has been paid to it compared with that paid to research data. Recently, there has been an increase in efforts to acknowledge and highlight the importance of software in research activities.

Structured metadata from platforms like \textit{bio.tools}, \textit{Bioconductor}, and \textit{Galaxy ToolShed} offers valuable insights into research software in the Life Sciences. Although originally intended to support discovery and integration, this metadata can be repurposed for large-scale analysis of software practices. However, its quality and completeness vary across platforms, reflecting diverse documentation practices.

To gain a comprehensive view of software development and sustainability, consolidating this metadata is necessary—but requires robust mechanisms to address its heterogeneity and scale.

This article presents an evaluation of instruction-tuned large language models for the task of software metadata identity resolution—a critical step in assembling a cohesive collection of research software. Such a collection is the reference component for the \textit{Software Observatory} at \textit{OpenEBench}, a platform that aggregates metadata to monitor the ``FAIRness'' of research software in the Life Sciences.

We benchmarked multiple models against a human-annotated gold standard, examined their behavior on ambiguous cases, and introduced an agreement-based proxy for high-confidence automated decisions. The proxy achieved high precision and statistical robustness, while also highlighting the limitations of current models and the broader challenges of automating semantic judgment in FAIR-aligned software metadata across registries and repositories.
\end{abstract}

\section{Introduction}

As scientific software becomes increasingly central to research activities, metadata describing these tools is scattered across a growing number of registries, repositories, package managers, and publication platforms. These platforms differ in scope, stability, and longevity—some emerge or disappear over time—leading to duplicate or outdated records for the same software across different sources.

The resulting discrepancies create ambiguity when integrating metadata: records may share names, developers, or even source code repositories but differ in function, domain, or completeness, making it difficult to determine whether individual records refer to the same software tool.

This ambiguity has practical consequences. Without identity resolution, it becomes harder to compute quality indicators, trace software usage, or enable consistent referencing. It also has a direct impact on identifying the contributions of software developers and, therefore, crediting such contributions beyond traditional peer-reviewed publications.

Traditional approaches to identity resolution have relied on rule-based systems or string similarity measures~\cite{binetteAlmostAllEntity2022, liRuleBasedMethodEntity2015, wellnerAdaptiveStringSimilarity2005, elliottSurveyAuthorName2010}. While these techniques offer computational efficiency, they often fail in the presence of sparse, inconsistent, or conflicting metadata, which is common in scientific registries where metadata may be auto-generated, partially maintained, or translated across formats. Therefore, more advanced mechanisms are needed to disambiguate software metadata.

Recent advances in large language models (LLMs) have opened new avenues for semantic classification, including entity linking and record disambiguation. Instruction-tuned models such as GPT-3~\cite{brownLanguageModelsAre2020}, T5~\cite{raffelExploringLimitsTransfer2019}, and FLAN~\cite{chungScalingInstructionFinetunedLanguage2022}, as well as more recent open-weight models like Mistral~\cite{jiangMistral7B2023}, Llama~\cite{touvronLLaMAOpenEfficient2023}, and Mixtral~\cite{jiangMixtralExperts2024}, have demonstrated strong capabilities in tasks requiring natural language understanding and reasoning~\cite{leAreLargeLanguage2023, murthyEvaluatingInstructionfollowingAbilities2024}. These models can follow structured prompts and perform classification with minimal supervision, making them attractive candidates for automating metadata integration in open research infrastructures.

In the context of the \textit{OpenEBench Software Observatory}\footnote{\href{https://openebench.bsc.es/observatory}{https://openebench.bsc.es/observatory}}, the use of LLMs offers the opportunity to enhance the process of software identity resolution. This step is crucial for all downstream analysis and observations.

While powerful models exist, integrating them into operational workflows requires balancing predictive quality with inference speed and confidence estimation. The current collection of software records available at the OpenEBench Software Observatory contains around 45{,}000 unique research software records, with the number of conflicts ranging between 500 and 3{,}000 cases, depending on pre-filtering assumptions. Despite representing only a small proportion (1--6\%) of the total number, these cases represent the most time-consuming and impactful resolution challenges. Addressing them is therefore essential for reliable metadata integration.

This study provided a controlled setting to benchmark LLMs for disambiguating metadata-based software identities, before scaling to more complex sources such as scientific literature. We offer an interpretable alternative with high practical reproducibility that complements existing rule-based approaches by evaluating LLM-based methods against a curated gold standard. 

Focusing on high-difficulty cases, we assessed the ability of LLMs to outperform heuristics, how closely they align with human judgment, and what trade-offs they introduce in terms of accuracy, annotation effort, and scalability. The result is a reusable foundation for integrating semantic resolution of software metadata within the OpenEBench Research Software Observatory.

% ----- Methods ------
\section{Methods} 
We implemented an evaluation pipeline to assess the performance of different LLMs in disambiguating software metadata records. Each LLM was compared to a human-annotated gold standard using classification metrics, error analysis on the so-called “hard” cases, and model agreement as a proxy for prediction confidence.

\subsection{Task Definition}
\label{subsec:task_definition}

The software identity resolution task was framed as a three-way classification problem: determining whether a pair of metadata records with the same name refers to the same software, refers to different software, or whether it is unclear due to insufficient information.

Each record included fields such as \texttt{name}, \texttt{description}, \texttt{repository URL}, \texttt{webpage URL}, \texttt{publication}, and \texttt{authors}, \texttt{developers} or \texttt{maintainers}. Additionally, the content of the referenced URLs was provided (see \nameref{subsec:prompting}), as reviewing the software’s website and repository is one of the primary methods a human would use to make a resolution decision.

% -------------- Gold Standard Construction ----------------

\subsection{Gold Standard Construction}
\label{subsec:gold_standard}

\paragraph{Dataset Selection and Sampling.}

To evaluate model performance, a representative set of 100 ambiguous software metadata cases was randomly sampled and manually annotated from a pool of 555 conflicting pairs identified when applying traditional methods.

The metadata was originally collected from various sources and homogenized after extraction to enable consistent comparisons. Ambiguous cases were identified as those where metadata records with the same name linked to different URLs (e.g., project websites or registries), or where records with different names pointed to the same URL (excluding source code repositories, which are considered a strong indicator of shared identity).

The initial pool of conflicting pairs was obtained after applying a series of assumptions to automatically resolve less ambiguous cases (e.g., considering records with matching names and non-repository URLs as the same software). Each remaining pair was identified as a potential identity conflict and assigned a unique identifier.

\paragraph{Annotation Process.}

Each record pair was annotated by a single human annotator with a verdict indicating whether the two records referred to the same software, different software, or were unclear.

The \textit{unclear} label was used when the annotator could not decide, typically due to broken or missing URLs combined with vague or generic metadata. Each confirmed case also received a confidence flag (\textit{low}, \textit{medium}, or \textit{high}) to support stratified performance analysis.

The annotator had access to the full content of the associated URLs, which could be reviewed in order to make informed judgments. Additionally, a brief rationale (one or two sentences) was noted for each verdict to provide transparency and traceability.

\paragraph{Annotation Metadata and Effort Tracking.}

All annotated cases were organized into a spreadsheet for readability, and the time dedicated to the annotation process was measured to assess the human effort involved.

\paragraph{Sampling Constraints and Class Imbalance.}

The final gold standard is not class-balanced. In practice, ambiguous metadata records referring to the same software are considerably more common than genuinely different records that share the same name.

We considered two alternatives to mitigate this imbalance: (1) continuing manual annotation until at least 50 cases were identified, or (2) fabricating negative examples by pairing unrelated records and modifying their names to create artificial naming conflicts.

The first option was prohibitively time-consuming, while the second would have required generating synthetic URL content aligned with the fabricated names—a complex task for both human annotators and LLMs alike.

Given these challenges, we chose to sample exclusively from real ambiguous cases and to accept the resulting class imbalance.

% ---------------- Models and Inference ----------------

\subsection{Models and Inference}
\label{subsec:models}

We benchmarked diverse instruction-tuned LLMs to evaluate their capacity for software identity resolution. Our selection criteria focused on three key dimensions: (1) diversity in model architecture and size, (2) openness and accessibility for future deployment, and (3) relevance to current trends in language model development.

\paragraph{Model Selection.}
The model pool included a range of open-weight LLMs, from compact 7B parameter architectures to larger sparse mixture-of-experts (SMoE) configurations. Open LLMs were prioritized for their transparency, accessibility, and alignment with the principles of FAIR research infrastructures. With the exception of one base model (Ministral 8B), all models used in the study were instruction-tuned, even if referred to by their short names for brevity (see Table~\ref{tab:technical} for a detailed technical overview). The inclusion of a non-instruction-tuned model also allows for an indirect view of the impact of instruction tuning on disambiguation performance. A single proprietary LLM, OpenAI GPT-4o, was also included to serve as a performance reference and to contextualize the results under optimal and closed-source conditions.

\paragraph{Inference Setup.}
All model inferences were conducted using publicly available inference APIs, specifically the Hugging Face Inference API \cite{HuggingFaceInference} and OpenRouter\cite{OpenRouterAPI}. This approach enabled a uniform and scalable benchmarking setup without the complexity of local deployment, which was considered out of scope for the benchmark phase. Inference was handled programmatically using consistent, chat-style prompt formatting across all LLMs.

All model outputs were generated using the same decoding parameters: \texttt{temperature=0.2}, \texttt{top\_p=0.95}, and \texttt{max\_new\_tokens=512}. These settings encourage focused but slightly variable outputs, which can improve response quality in open-ended tasks like metadata disambiguation. The \texttt{return\_full\_text} flag was set to \texttt{False} to isolate the generated response. Random seeds were not explicitly set, and for some model interfaces, seed control may not have been available; therefore, outputs are not strictly reproducible, although model behavior was observed to be stable across runs. The exact model identifiers used via each inference API are listed in Table~S1. These correspond to the specific model aliases exhibited by the Hugging Face Inference API or OpenRouter at the time of evaluation. Note that some aliases may map to custom checkpoints or provider-specific variants and may change over time.

\paragraph{Prompt Standardization.}
To ensure comparability, a single prompt format was used for all LLMs. Metadata records and content were passed as structured Markdown blocks within chat messages (see \nameref{subsec:prompting}), and the output was expected to be in a JSON format containing a verdict, confidence, and explanation. Outputs were parsed and validated automatically; failures to conform to the expected format resulted in the record being marked as skipped.

Table~\ref{tab:technical} outlines technical specifications including parameter count, model architecture, context window size, release date, and whether the model has been instruction-tuned.

\paragraph{Inference Timing.}
We recorded the total time elapsed for each model and metadata pair from request submission to response receipt, capturing end-to-end latency. When available, internal latency metrics reported by the API (e.g., from OpenRouter) were also logged, allowing us to distinguish between model processing time and network overhead.

These timing measurements were used to assess each LLM’s operational cost and compare automated and manual annotation effort. They also informed discussions of model responsiveness and feasibility for integration into Extract, Transform, and Load (ETL) workflows.

\begin{table}[ht]
\begin{tabularx}{0.5\textwidth}{llX}
\hline
\textbf{Model} & \textbf{License} & \textbf{Open Weights}  \\ 
\hline
Mixtral 8x7B       & Apache 2.0      & Yes           \\
Mixtral 8x22B        & Apache 2.0      & Yes          \\
Mistral 7B       & Apache 2.0      & Yes           \\
Ministral 8B          & Apache 2.0      & Yes          \\
Llama 4 Scout     & Custom (Meta)   & Yes           \\
OpenChat           & Apache 2.0      & Yes          \\
Llama 3.3          & Custom (Meta)   & Yes           \\
GPT-4o             & Proprietary     & No           \\ 
\hline
\end{tabularx}
\caption{Openness and accessibility of evaluated models.}
\label{tab:accessibility}
\end{table}

\begin{table*}[ht]
\centering
\begin{tabularx}{\textwidth}{lXXll}
\hline
\textbf{Model} & \textbf{Instruction-tuned} & \textbf{Size / Type} & \textbf{Context Window} & \textbf{Release} \\
\hline
Mixtral 8x7B       & Yes                 & 46.7B (SMoE)          & 32K tokens              & 2024 \\
Mixtral 22B        & Yes                 & 141B (SMoE)           & 32K tokens              & 2024 \\
Mistral 7B       & Yes                       & 7.3B                  & 32K tokens              & 2024 \\
Ministral 8B          & No                       & 8B                    & 32K tokens              & 2024 \\
Llama 4 Scout      & Yes                & 17B                   & 128K tokens             & 2025 \\
OpenChat           & Yes                 & 7B                    & 32K tokens              & 2024 \\
Llama 3.3          & Yes                & 70B                   & 128K tokens             & 2024 \\
GPT-4o             & Yes & Not disclosed & 128K tokens  & 2024 \\
\hline
\end{tabularx}
\caption{ Technical characteristics of evaluated models.}
\label{tab:technical}
\end{table*}

% ---------------- Prompting ----------------

\subsection{Prompting}
\label{subsec:prompting}

\paragraph{Prompt Format and Structure.}
All LLMs were prompted using a standardized instruction template followed by structured content blocks representing metadata and contextual information for each record (see Listing S1 for the full prompt). Prompts were formatted in chat style, as required by both the Hugging Face Inference API and OpenRouter API, with one message per metadata record followed by a final instruction message.

The prompt was structured to align with the expected behaviour of instruction-tuned LLMs~\cite{ouyangTrainingLanguageModels2022a}. For consistency across models and interfaces, all messages, including the initial task description, were passed as \texttt{"user"} messages; no \texttt{"system"} message was used. This design was chosen to ensure portability across providers and was found to yield reliable behaviour in practice. Each prompt consisted of:

\begin{itemize}
    \item An initial instruction message describing the identity resolution task and the expected output format (\texttt{verdict}, \texttt{confidence}, and \texttt{explanation}).
    \item A user message containing the metadata for the first record, including fields such as name, description, repository URL, webpage, authors, and publications.
    \item A second user message with the metadata of the second record.
    \item Additional user messages containing the cleaned content of associated URLs (e.g., repository README or project website), one per record.
    \item A final instruction message reminding the model of the required output format and indicating that it can now begin reasoning and respond.
\end{itemize}

\paragraph{Metadata and Context Handling.}
Metadata records were rendered as nested dictionaries within Markdown code blocks to preserve structure and improve interpretability. All fields were included where available. URL content was placed after the metadata and extracted using a combination of tools:

\begin{itemize}
    \item \textbf{Generic websites:} Scraped using Playwright\cite{MicrosoftPlaywrightpython2025}, with non-relevant HTML elements removed via BeautifulSoup\cite{leonardrichardsonBeautifulsoup4ScreenscrapingLibrary}. The result was converted to Markdown, preserving basic structure and links.
    \item \textbf{Structured sources:} For GitHub, GitLab, Bitbucket, and PyPI, dedicated extractors using their respective APIs were implemented. For SourceForge, tailored HTML parsers were created to extract meaningful sections using Playwright and class-based tag filtering.
\end{itemize}

\paragraph{Prompt Design Strategy.}
Token length was monitored during prompt construction, but no truncation was required. To prevent LLMs from overfitting to specific examples, we avoided few-shot prompting. The use of a concrete output sample included in the prompt was discarded due to excessive mimicry in model responses, and instead a more abstract reminder for such output structure was used.

The prompt was iteratively refined using Mistral 7B as a reference model, as smaller LLMs tend to be more sensitive to prompt design. Refinements included:

\begin{itemize}
    \item Explicit guidance on reasoning strategy.
    \item A reminder that the records may share the same name but name similarity alone is not a reliable resolution signal.
    \item Stricter formatting cues for the expected JSON output.
\end{itemize}

\paragraph{Response Validation.}
Model responses were parsed automatically. If the response failed to produce a valid JSON object or omitted mandatory fields, the record was marked as \textit{skipped}, though raw outputs were retained for manual inspection and error analysis.

\subsection{Evaluation Metrics}

To evaluate model performance on the software identity resolution task, we combined standard classification metrics with focused error analyses designed to capture the semantic and practical challenges specific to this benchmark. Our goal was not only to assess overall predictive accuracy, but also to understand how closely the LLMs’ behavior aligns with human reasoning in cases requiring nuanced interpretation.

\paragraph{Core Metrics.}
We calculated the following core metrics: accuracy, macro-averaged F1-score, macro-averaged precision, and macro-averaged recall. Accuracy provided a straightforward measure of overall correctness. In macro-averaging, the metric (e.g., precision, recall, or F1) is first computed independently for each class and then averaged across classes, giving equal weight to each regardless of how often it appears in the data. For example, to compute the macro F1-score, we first calculated the F1-score~\cite{rijsbergenInformationRetrieval1979, christenReviewFMeasureIts2023} for each of the three target labels (“same”, “different”, and “unclear”), and then took the unweighted average of these three values. This approach ensures a balanced assessment even in the presence of class imbalance. 

While macro-F1 is widely used, it has also been criticized as a non-representational measure\footnote{The F1-score is not representational in the strict sense of measurement theory. Rather than reflecting a consistent empirical quantity, it functions as a heuristic that combines precision and recall into a single value to summarize their trade-off. This makes it useful for practical evaluation, especially in imbalanced settings, but also limits its interpretability and the validity of operations such as averaging across classes or datasets.}, since averaging harmonic means does not preserve meaningful mathematical properties~\cite{powersWhatFmeasureDoesnt2019}. We include it here as a pragmatic indicator of balanced class-level performance.

To address the limitations of aggregate metrics—such as their tendency to hide how well the model performs on each individual label and whether it favors precision over recall—we also computed per-class precision and recall, as well as confusion matrices for each model. These more detailed evaluations provide a clearer picture of strengths and weaknesses across the different classes. For all reported metrics except confusion matrices, we calculated 95\% confidence intervals using bootstrap resampling with 1,000 iterations, sampling from the set of cases that were resolved by each model or proxy.

\paragraph{Difficulty-Sensitive Evaluation.}
To probe model robustness and alignment with human reasoning, we stratified the evaluation by difficulty. Each case in the gold standard was annotated by a single annotator with a verdict and a confidence rating. For evaluation purposes, we grouped cases into either “hard” or “easy” according to the degree of identity resolution difficulty. Hard cases were those where the annotator either selected the “unclear” label or marked their verdicts low-confidence. Easy cases were those for which the annotator expressed medium or high confidence in assigning the “same” or “different” label. 

While three confidence levels were recorded during annotation, this binary grouping offers a more conservative and robust distinction, given the subjectivity inherent in single-annotator assessments. This distinction allowed us to examine whether LLMs struggle in situations that challenge humans. Their performance on hard cases served as an indicator for semantic reliability and alignment with human evaluative strategies—an important trait of automated decisions. 

For each model, we computed the error rate separately on the easy and hard subsets, defined as the proportion of incorrect predictions within each group. To assess the uncertainty around these estimates, we computed 95\% confidence intervals using stratified bootstrap resampling (1,000 iterations). We then tested for a statistically significant difference between error rates on hard and easy cases using a bootstrap hypothesis test on the difference in means. This allowed us to identify LLMs that were significantly more likely to make errors on hard cases, while accounting for differences in sample size between subsets.

\subsection{Agreement-Based Decision Proxy}

In addition to evaluating individual model predictions, we introduced an agreement-based proxy for high-confidence automated resolution. A prediction was accepted automatically only when all top-performing LLMs agreed; cases with disagreement were deferred to human review. This proxy was evaluated using the same metrics as the individual LLMs—accuracy, macro-averaged F1-score, macro-averaged precision, and macro-averaged recall—and its performance was reported in Section~\ref{subsec:proxy-evaluation}.

\begin{table}[ht]
\begin{tabularx}{0.5\textwidth}{ll}
\hline
\textbf{Proxy} & \textbf{Composition}  \\ 
\hline
 Proxy I & Llama 4 Scout + Mixtral 8x22B \\
Proxy II & Mistral 7B + Mixtral 8x22B \\
Proxy III & Mixtral 8x7B + Mixtral 8x22B \\
Proxy IV & Mistral 7B + Mixtral 8x7B \\
Proxy V & Llama 4 Scout + Mixtral 8x7B \\
\hline
\end{tabularx}
\caption{Composition of the agreement proxies we assessed.}
\label{tab:proxies_composition}
\end{table}

We selected the three best-performing LLMs overall and included Mixtral-8x22B due to its human-like behavior of performing significantly better on easy cases than on hard ones, which makes it a useful indicator of ambiguity (Table~\ref{tab:proxies_composition}). Other combinations were not considered, as they either involved lower-performing LLMs or showed inferior performance as proxies, reducing their practical utility.

% ----- Results ------
\section{Results} 
\label{sec:results}

\subsection{Gold Standard Distribution}
\begin{figure*}[h!]
    \centering
    \includegraphics[width=\textwidth]{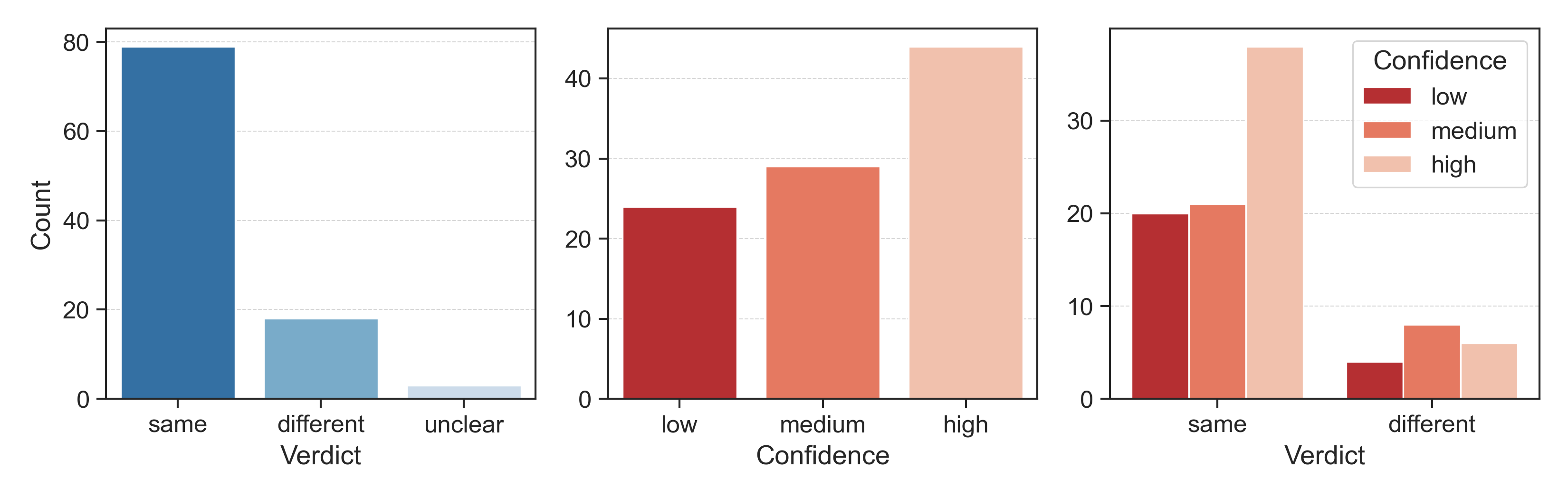}
    \caption{Distributions of human annotations in the gold standard. Left: class distribution across the three possible verdicts (\texttt{same}, \texttt{different}, \texttt{unclear}). Center: distribution of annotator confidence levels across all annotated cases. Right: joint distribution of verdict and confidence, restricted to entries labeled as \texttt{same} or \texttt{different}. \texttt{Unclear} cases were excluded from confidence scoring.}
    \label{fig:standard_distributions}
\end{figure*}

The gold standard comprised 100 software metadata cases selected for their ambiguity. As shown in Figure~\ref{fig:standard_distributions}, the majority of records were labelled as “same”, reflecting the observation that duplicated software names are more frequently associated with the same entity than with unrelated projects. Fewer records were labelled “different”, and an even smaller subset received the “unclear” label, typically due to missing URLs or vague metadata. 

Annotator confidence followed a skewed distribution, with most cases receiving a high rating and a smaller number labelled low or medium. The joint distribution (Figure~\ref{fig:standard_distributions}, right panel) revealed that the “same” verdicts tended to be rated with higher confidence, while “different” cases showed more variability. Unclear cases were not assigned a confidence rating, since no actual decision was made and confidence scores apply only to resolved cases.

\subsection{Language Model Performance}

\begin{figure*}[h!]
    \centering
    \includegraphics[width=\textwidth]{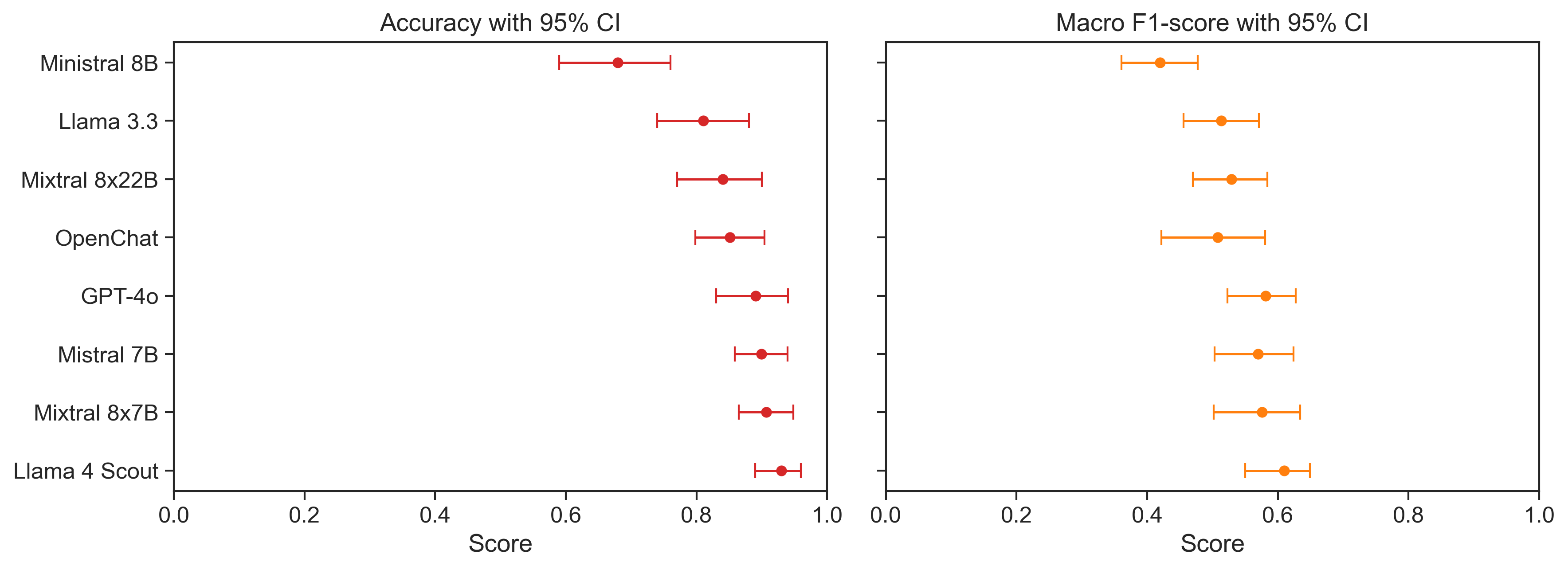}
    \caption{Multiclass evaluation metrics by model. Each bar shows accuracy and macro-F1 for a given language model.}
    \label{fig:model_metrics}
\end{figure*}

Figure~\ref{fig:model_metrics} summarizes model performance in terms of accuracy and macro F1-score. All metrics were computed over non-skipped records only (see Table S2 for the number of cases resolved by each LLM). 

Top-performing LLMs, including Llama 4 Scout, GPT-4o, Mistral 7B, and Mixtral 8x7B, achieved bootstrap-estimated accuracies with means above 0.89. The lower bounds of their 95\% confidence intervals remained above 0.83, indicating consistently high accuracy across resampled subsets. In contrast, macro F1-scores for these LLMs ranged from 0.57 to 0.61, with wider confidence intervals. This discrepancy reflects the influence of class imbalance and the systematic failure across all LLMs to correctly classify the three “unclear” cases, which consistently yielded zero precision and recall. As a result, macro-average performance—especially recall—is penalized, highlighting the challenge these LLMs face in capturing rare or ambiguous cases. In contrast, other LLMs such as OpenChat, Llama 3.3, and Ministral 8B showed weaker performance, particularly in macro F1, suggesting difficulty handling all class types. 

In addition to the core metrics presented in the main text, we report per-class precision and recall in the supplementary material (Table S3). These reveal complementary strengths and weaknesses across LLMs. Llama 4 Scout exhibited the highest macro precision and recall, with perfect recall on the cases labelled as “same” and perfect precision on cases labelled as “different”. Mixtral 8x7B showed the highest precision for the “same” cases. GPT-4o also performed well, achieving a balanced profile across both the “same” and “different” classes. As noted above, all LLMs failed to correctly identify “unclear” cases, indicating that reporting semantic uncertainty remains challenging and may require separate handling in production settings. Detailed confusion matrices for all LLMs are presented in Figure S1.

\begin{figure*}[h!]
    \centering
    \includegraphics[width=\textwidth]{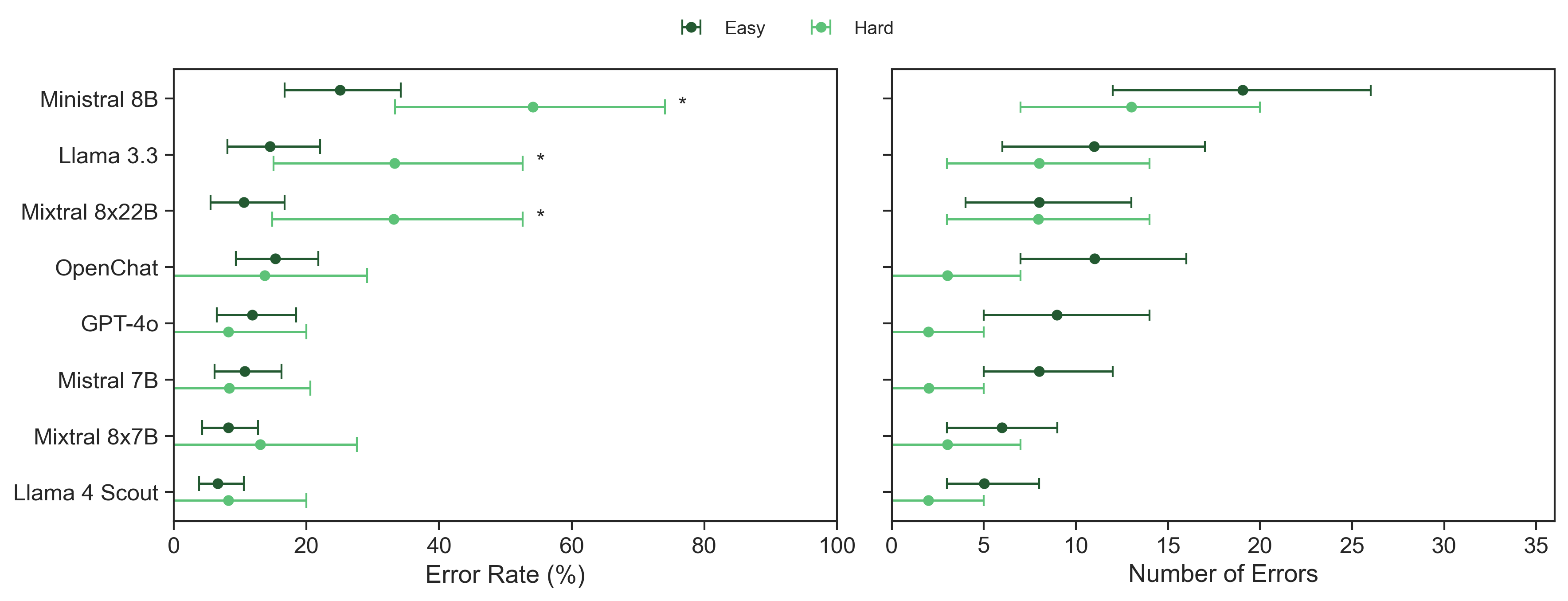}
    \caption{Comparison of model performance on “hard” vs. “easy” disambiguation cases. The left panel shows error rates with 95\% bootstrapped confidence intervals; each dot represents the mean error rate for a given case type (easy or hard). The right panel displays the corresponding number of errors on each case type. Asterisks (*) indicate LLMs for which the error rate on hard cases was significantly higher than on easy cases ($p < 0.05$, bootstrap test).}
    \label{fig:error_rates}
\end{figure*}

Error rates were generally higher on cases labeled as “hard” by human annotators, but in most LLMs the difference compared to “easy” cases was small and not statistically significant (Figure~\ref{fig:error_rates}). Confidence intervals for hard cases tended to be wider, reflecting their smaller number in the dataset. Only a subset of LLMs—such as Mixtral 8x22B, Ministral 8B, and LLaMA 3.3—showed a significant increase in error rate on hard cases ($p < 0.05$, bootstrap test), and these were also among the LLMs with the highest overall error rates. This suggests that sensitivity to disambiguation difficulty is more pronounced in less accurate LLMs, while stronger LLMs maintained more consistent performance across difficulty levels.

\subsection{Agreement-Based Proxy Evaluation}
\label{subsec:proxy-evaluation}

\begin{figure*}[h!]
    \centering
    \includegraphics[width=\textwidth]{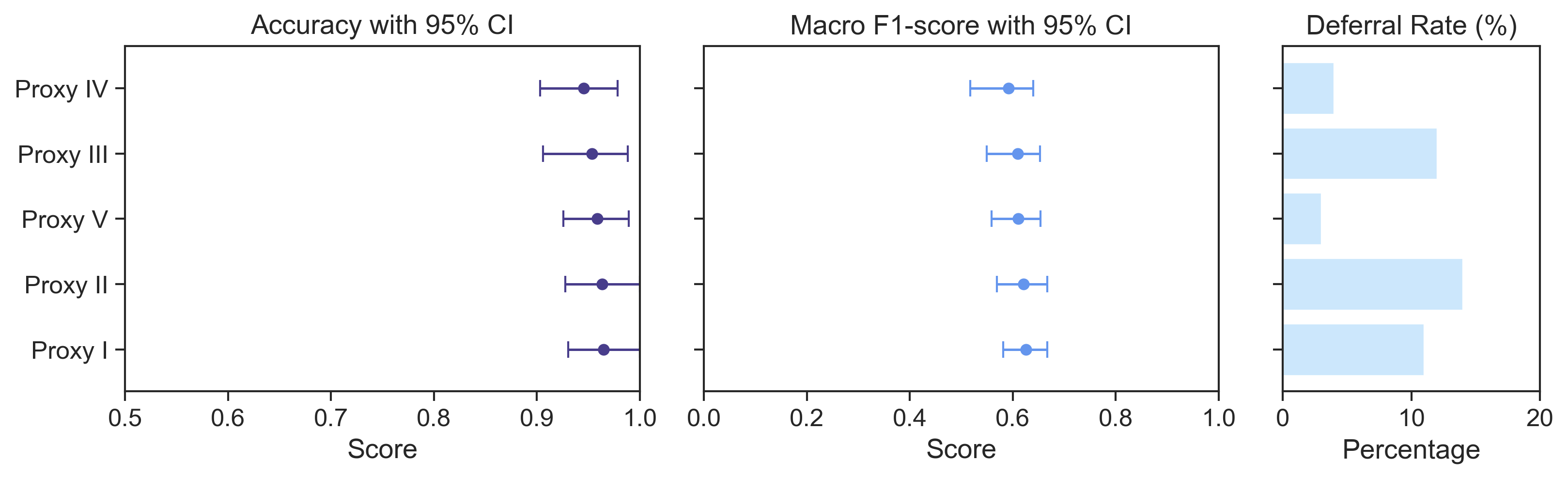}
    \caption{Performance and coverage of agreement-based proxies. Each point shows the accuracy and macro F1-score of a proxy with 95\% bootstrap confidence intervals. The rightmost panel shows the percentage of cases deferred because there was a disagreement between the LLMs. All metrics are computed only over non-deferred case}
    \label{fig:proxies_metrics}
\end{figure*}

Among the proxies (Figure \ref{fig:proxies_metrics}), Proxy I (LLaMA 4 Scout + Mixtral 8x22B) achieved the strongest overall performance, with the highest accuracy (0.965; 95\% CI: 0.930-1.000) and macro F1-score (0.626; 95\% CI: 0.581-0.667). It also yielded perfect precision on  the “different” class and high (0.858; 95\% CI: 0.920-1.000) on the “same” class.  It also achieved the highest recall on the “different” (0.822; 95\% CI 0.647-1.000) class, while issuing verdicts for 86 out of 100 benchmark cases. This makes Proxy I the most precise and semantically reliable configuration tested. 
In contrast, Proxy V (LLaMA 4 Scout + Mixtral 8x7B) offered slightly lower performance but significantly higher coverage, producing confident verdicts for 94 out of 100 cases. With an accuracy of 0.958 (95\% CI: 0.926-0.989) and a macro F1-score of 0.611 (95\% CI: 0.559-0.654), Proxy V delivered a strong balance between reliability and coverage. Like Proxy I, it also attained high precision for both “same” and “different” classes, with a slightly reduced recall on “different” (0.756; 95\% CI: 0.562-0.938). 
The remaining proxies illustrated how different pairing strategies —whether based on difficulty sensitivity, architecture similarity, or model size— affect the precision-coverage trade-off. Proxy IV, for instance, combined two smaller LLMs (Mistral 7B + Mixtral 8x7B) and achieved the highest number of agreement cases (93) with commendable precision, although its recall on “different” cases was the lowest of all proxies. 
Taken together, these results suggest that the model  agreement can be a highly effective proxy for prediction confidence, and that the pairing of complementary LLMs —one high-performing and one conservative— offers a robust basis for selective automation in metadata resolution tasks.

\subsection{Annotation Time: Human vs Model}

\begin{figure}[h!]
    \centering
    \includegraphics[width=\columnwidth]{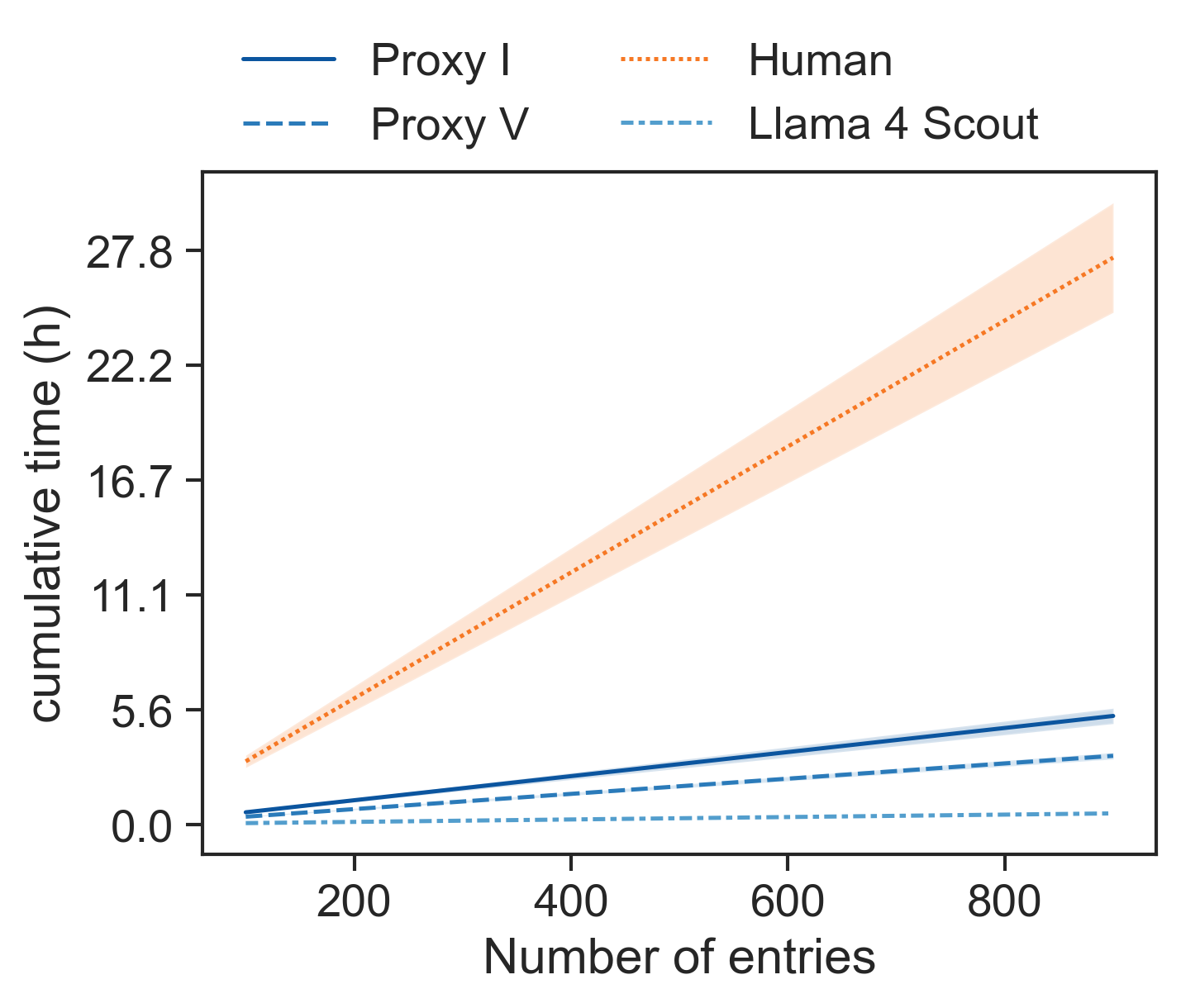}
    \caption{Extrapolated cumulative annotation time as a function of dataset size for humans, Llama 4 Scout, and proxy methods. All time estimates are extrapolated from a shared gold standard of 100 annotated records. The human curve is based on measured total annotation time across those 100 cases, with a 95\% confidence interval shown as a shaded area. Llama 4 Scout time is derived from average end-to-end latency per record across the same set. Proxy I and Proxy V simulate agreement-based workflows, combining Llama 4 Scout with Mixtral 8x22B and Mixtral 8x7B, respectively, with fallback to human annotation in 15\% (Proxy I) and 6\% (Proxy V) of cases. No full-scale annotation was performed beyond the initial 100; curves represent linear projections.
}
    \label{fig:annotation_time}
\end{figure}

The comparison in Figure~\ref{fig:annotation_time} shows that LLMs completed the annotation task significantly faster than humans. While the absolute latency per model varies depending on the inference provider and network routing, the overall gap between human and automated annotation times remains substantial. The figure also includes two proxy strategies—Proxy I and Proxy V—which combined model agreement with fallback to human annotation in 14\% and 6\% of cases, respectively. These proxy methods offer a balance between automation and human oversight while still yielding considerable reductions in annotation time compared to fully manual curation.

% ----- Discussion -----
\section{Discussion}

Intensive data-driven research activities rely heavily on software. However, little attention has been paid to its quality and sustainability. Cataloguing software is a necessary step to understand current practices and propose targeted actions to improve its quality and, therefore, contribute towards its sustainability. However, metadata describing software tends to be scattered across multiple sources and is often incomplete or outdated. Thus, metadata-based identity resolution mechanisms are needed as part of the cataloguing efforts.

The results of this study highlight the potential of instruction-tuned LLMs to support accurate and scalable identity resolution of software metadata. Several LLMs strongly aligned with human judgment, achieving accuracy above 89\% and macro F1-scores approaching or exceeding 0.60. These results suggest that even in tasks requiring semantic interpretation of sparse or conflicting metadata, modern LLMs can perform reliably well with minimal supervision.

Performance varied across LLMs and evaluated classes. While larger LLMs and those trained with reinforcement-based alignment techniques (e.g., GPT-4o, Llama 4 Scout) often performed well, the overall difference in performance between large and small LLMs was less pronounced than expected. Surprisingly, several smaller open-weight LLMs (e.g., Mistral 7B, Mixtral 8x7B) approached or matched the accuracy and agreement scores of significantly larger LLMs. This suggests that, for this classification task, model size alone does not guarantee superior performance and that smaller LLMs may already encode sufficient reasoning capability when given structured input and clear instructions. It also highlights the potential gains to be made by explicitly tailoring prompts for smaller LLMs, which could further narrow the performance gap in practice.

In contrast, the lowest performance was observed for Ministral 8B, the only non-instruction-tuned model in the benchmark. This is consistent with the central role that instruction tuning plays in enabling models to effectively follow task prompts and produce structured, goal-directed outputs.

Performance differences between “easy” and “hard” cases—based on human annotation—were generally modest across models, and in most cases not statistically significant. This suggests that stronger LLMs were able to maintain robustness even when facing more ambiguous or challenging examples. However, three models stood out for showing a significantly higher error rate on hard cases: Mixtral 8x22B, LLaMA 3.3, and Ministral 8B. Among them, Ministral 8B is particularly notable. Despite being the lowest-performing model overall—likely due to being the only non-instruction-tuned model—it was also the one whose performance most closely mirrored human-perceived difficulty. This alignment suggests a degree of interpretability that is often absent in more capable but opaque models. Given its compact size and transparent behavior, it may be worthwhile to explore the potential of an instruction-tuned version of Ministral 8B, which could improve its effectiveness while preserving its apparent sensitivity to task complexity.

The results also demonstrate that model agreement can serve as a highly effective mechanism for automating software identity resolution with high precision. The strongest proxy overall, Proxy~I (Llama 4 Scout + Mixtral 8x22B), combined a high-performing model with a difficulty-sensitive model. Mixtral 8x22B was one of the few LLMs for which the error rates were statistically correlated with human-labeled difficulty, suggesting that it was sensitive to semantic ambiguity. In contrast, Llama 4 Scout showed overall high performance and confidence, with no such correlation. This combination of pairing difficulty-awareness with decisive accuracy proved especially effective. Proxy~I achieved the highest accuracy (0.965), macro F1-score (0.941), and Cohen’s Kappa (0.882), while maintaining perfect precision on both “same” and “different” cases. Its superior recall on different records (0.824) further underscores its value in reducing false matches, a key requirement for robust metadata integration.

However, Proxy~V (Llama 4 Scout + Mixtral 8x7B) offers a compelling alternative for large-scale deployments. It returned verdicts for 94 out of 100 cases, the highest among the high-performing proxies, while still achieving excellent accuracy (0.957) and perfect precision. Although slightly less conservative than Proxy~I, its broader coverage makes it particularly attractive for operational settings that aim to automate as many cases as possible while maintaining trustworthiness.

The results also highlighted how proxy composition influences behavior. Pairing LLMs with complementary behavior and architecture—such as two instruction-tuned LLMs of different types, like a dense decoder (Llama 4 Scout) and a sparse mixture-of-experts model (Mixtral)—appears to produce proxies that are both selective and semantically precise.

Importantly, all proxies failed to identify any unclear cases. This limitation points to a broader challenge in automating ambiguity recognition and underscores the need for alternative strategies such as abstention-aware prompting, uncertainty modeling, or explicit deferral mechanisms.

Overall, these findings suggest that agreement-based proxies—particularly when carefully constructed from complementary LLMs—can deliver near-human precision on a substantial portion of cases. This approach offers a scalable and reliable mechanism for automated metadata integration in FAIR-aligned software observatories.

Although LLMs are not instantaneous, their annotation speed is orders of magnitude faster than manual curation, even when accounting for network latency and API overhead. This efficiency is critical for assembling and scaling up the OpenEBench Software Observatory’s software collection. However, such speed comes with computational costs. LLMs like Llama 4 Scout, and especially Mixtral 8x22B, require substantial resources, including high-memory GPUs and parallel infrastructure, which can limit accessibility and raise both environmental and economic concerns. To mitigate this, a pragmatic approach is to reserve LLM inference for cases that remain unresolved after applying lightweight heuristics or string-matching rules. This triage strategy reduces unnecessary model calls while preserving annotation quality. Proxy strategies like Proxy~I and Proxy~V further support the feasibility of hybrid human-AI workflows by maintaining annotation quality through selective deferral, without sacrificing scalability. These results highlight the practical viability of integrating LLMs into metadata workflows, enabling both speed and precision at scale.

Despite these promising results, there remains room for improvement, both in model prompting and input preparation. Some prediction errors appear to stem from how metadata is structured or presented. Future iterations could explore reformatting metadata by replacing code-style nested dictionaries with natural-language-like field lists (e.g., \texttt{Name: diamond, Version: 1.03}), improving readability and alignment with instruction-tuned model expectations. Systematically removing redundant or irrelevant fields may also reduce noise and cognitive load. On the content side, website readability could be enhanced by integrating tools like Postlight Parser\cite{PostlightParserExtract} alongside Playwright and BeautifulSoup, enabling better extraction of relevant information from semi-structured sources.

To further improve system performance and applicability, we also plan to refine prompts for both small and large LLMs, expand the annotated dataset through targeted manual labeling, and improve pre-processing workflows. We plan to fine-tune the LLMs using cases that are deferred by the agreement proxy and subsequently annotated by humans, enabling targeted improvements where the LLMs show poor performance.

In addition, involving multiple annotators and measuring inter-annotator agreement will be essential to better characterize task ambiguity and establish a more meaningful upper bound on achievable model performance.

%------ Conclusions ----
\section{Conclusion}

This study benchmarked instruction-tuned language models for the task of software identity resolution, a critical yet underexplored step in ensuring data quality within open research infrastructures. We evaluated several LLMs on a manually curated gold standard, assessed their behavior on ambiguous cases, and introduced a model-agreement-based proxy for high-confidence classification.

Our results showed that both large and small LLMs can achieve strong alignment with human annotations, with several open-weight LLMs performing competitively with proprietary alternatives. Notably, a simple agreement-based proxy yields both high precision and wide coverage, offering a practical mechanism for selective automation.

In addition to strong accuracy, the use of LLMs significantly reduces annotation time, suggesting that automated identity resolution is both feasible and scalable for the assembly of the software collection at the OpenEBench Software Observatory. The findings support the integration of LLMs into FAIR-aligned metadata workflows and highlight the viability of lightweight consensus mechanisms for high-confidence automation.

%\bibliography{identity_resolution}

\end{document}